\documentclass[12pt]{article}

\usepackage[utf8]{inputenc}
\usepackage[T1]{fontenc}
\usepackage{mathpazo}
\usepackage[final,expansion=false]{microtype}
\usepackage[letterpaper, margin=1in]{geometry}
\usepackage{setspace}
\usepackage{amsmath, amssymb, amsfonts, amsthm}
\usepackage{mathtools}
\usepackage{bm}

\theoremstyle{plain}
\newtheorem{assumption}{Assumption}
\newtheorem{theorem}{Theorem}

\newtheorem{proposition}{Proposition}

\theoremstyle{definition}

\newtheorem{remark}{Remark}

\newcommand{\diag}{\operatorname{diag}}
\newcommand{\one}{\mathbf{1}}
\newcommand{\ind}[1]{\mathbf{1}\{#1\}}
\newcommand{\bR}{\mathbb{R}}
\newcommand{\cS}{\mathcal{S}}
\newcommand{\cP}{\mathcal{P}}
\newcommand{\cN}{\mathcal{N}}

\usepackage{graphicx}
\usepackage[font={footnotesize},justification=justified]{caption}
\usepackage{booktabs}
\usepackage{adjustbox}

\usepackage{enumitem}

\usepackage[svgnames]{xcolor}
\usepackage[colorlinks=true, linkcolor=black, citecolor=black,
            urlcolor=blue]{hyperref}
\usepackage{natbib}
\usepackage[capitalize, noabbrev, nameinlink]{cleveref}
\crefname{assumption}{Assumption}{Assumptions}

\usepackage{tikz}
\usetikzlibrary{calc, positioning}
\usepackage{pgfplots}
\pgfplotsset{compat=1.18}
\usepgfplotslibrary{ternary}

\usetikzlibrary{arrows.meta}

\tikzset{declare function={
  sx(\a,\b)=(3*exp(\a)+6*exp(\b))/(1+exp(\a)+exp(\b));
  sy(\a,\b)=3*sqrt(3)*exp(\a)/(1+exp(\a)+exp(\b));
}}

\title{Compositional Difference-in-Differences\thanks{%
Corresponding address: oboussim@calpoly.edu, Department of Economics, The California Polytechnic State University, San Luis Obispo. I am grateful to Marc Henry, Andres Aradillas-Lopez, and Sung Jae Jun for their guidance on this project. I also thank Keisuke Hirano, Ismael Mourifie, Desire Kedagni, and Patrick Guggenberger, as well as all participants of the Penn State Econometrics Seminar, the Penn State Student Econometrics Workshop, the RESA Alumni Conference 2025, and the EconNect Africa JMC Symposium, for their valuable comments and suggestions. All remaining errors are my own.}}
\author{Onil \textsc{Boussim}}

\date{\today}

\begin{document}

\maketitle

\begin{abstract}
\noindent
Many causal questions involve outcomes distributed across mutually exclusive categories, such as votes by party or employment by status, where both category shares and the total quantity are of interest. This paper develops Compositional Difference-in-Differences (CoDiD), a framework for causal inference with categorical and compositional outcomes. Focusing first on shares, I introduce parallel trends in log-odds (PTLO), which requires the untreated log-odds of each category relative to a reference category to evolve in parallel across groups. I show that PTLO admits three equivalent interpretations: stability over time of the dependence structure between untreated categorical outcomes and group membership; parallel evolution of relative utilities in a random-utility model when shares arise from discrete choices; and parallel trajectories in the Aitchison geometry of the simplex. A stronger assumption, parallel growth in log-quantities, jointly identifies treatment effects on category shares and the total quantity. I also derive sharp bounds under a relaxation of the identifying assumption. Finally, I illustrate the framework by examining the effects of early voting on U.S. presidential elections.

\noindent\textbf{Keywords:} categorical outcomes, difference-in-differences, causal inference, compositional data, simplex, voting\par
\noindent\textbf{JEL codes:} C21, C25, D72.
\end{abstract}

\newpage

\section{Introduction}
\label{sec:intro}

Many causal questions in difference-in-differences (DiD) settings involve vectors of categorical quantities that together form a well-defined total.
These arise naturally when individual-level categorical outcomes are aggregated. In evaluating a minimum wage policy, individuals may be classified
as unemployed, part-time employed, or full-time employed, and the compositional vector counts how many fall into each group. In other cases, the
data consist directly of quantities across categories with no underlying individual identifiers, such as the number of votes received by each party in an election. Compositional vectors also appear when a single quantity is
decomposed into meaningful parts: electricity generation by source, GDP by
sector, or a budget allocated across expenditure categories. In all such cases,
the outcome is multidimensional, and both the shares and the total can matter
for the analysis.

Researchers typically focus on shares, and the canonical empirical strategy
analyzes each category in isolation, running a separate linear DiD on each
share. This practice has well-known defects. Imposing additive parallel
trends on shares is inconsistent with the simplex constraint and can generate
counterfactual shares outside $[0,1]$
\citep{lechner2011estimation,athey2006identification,wooldridge2023simple}.
Estimating each category separately also cannot, by construction, describe
how a policy reallocates mass across mutually exclusive alternatives.
Existing alternatives, geodesic DiD \citep{zhou2025geodesic},
transition-based methods \citep{ahn2025event}, and fractional response DiD
\citep{wooldridge2023simple}, address some of these issues, but their
counterfactuals are not generally consistent with a behavioral model of how
participants choose among alternatives, and they do not coherently deliver
identification of effects on totals.

In fact, in many applications, treatment effects on category shares and total quantities are jointly relevant. A voting reform may change both vote shares and turnout; a minimum wage may alter the composition and overall level of employment; and a carbon policy may affect both the energy mix and total electricity generation. Existing approaches typically estimate these effects separately, potentially relying on different or incompatible identifying assumptions. This paper develops Compositional Difference-in-Differences (CoDiD), a unified framework for causal inference on compositional outcomes and their associated totals. The framework proceeds in two steps.

First, I focus on the compositional margin and introduce parallel trends in log-odds (PTLO). This assumption requires that, absent treatment, the log-odds of each category relative to a reference category evolve in parallel across treated and comparison groups. Equivalently, the pairwise odds ratios between groups remain constant over time. Under PTLO, the treated group's counterfactual composition is identified by multiplying its baseline shares by the comparison group's category-specific share-growth factors and normalizing the resulting vector. This procedure yields a valid counterfactual composition in closed form, with strictly positive shares that sum to one. 

PTLO admits three complementary interpretations. First, a discrete-copula interpretation: since odds ratios characterize the margin-free association between categorical outcomes and group membership, PTLO requires this dependence structure to remain stable over time in the absence of treatment. Using the discrete-copula construction of \citet{geenens2020copula}, this restriction can be expressed as the stability of a standardized dependence table, while allowing the marginal distributions to be unconstrained. Second, a behavioral interpretation: under a multinomial-logit random-utility model with a common error scale, PTLO is equivalent to parallel changes in relative systematic utilities across groups. Thus, in the absence of treatment, the relative attractiveness of any pair of alternatives would evolve identically in treated and comparison populations. Third, a geometric interpretation: PTLO implies parallel trajectories in the Aitchison geometry of the simplex, where changes in composition are represented through log-ratios \citep{aitchison1982statistical,egozcue2003isometric}.

Second, I extend the framework to jointly identify treatment effects on category shares and total quantities. The key assumption is parallel growth in log-quantities, which requires untreated category-specific quantities to grow at the same proportional rates across groups. This stronger restriction implies PTLO for the compositional margin, preserving its dependence-based, geometric, and behavioral interpretations, while also identifying the counterfactual total. Importantly, it reveals a composition adjustment factor that accounts for aggregation bias in conventional log-total DiD. Even when category-specific growth rates are common across groups, differences in their initial compositions can generate different growth rates in aggregate quantities. The adjustment factor captures this discrepancy and provides a composition-consistent counterfactual for the total. Finally, I relax the identifying assumption using information from multiple pre-treatment periods, deriving partial-identification bounds on counterfactual outcomes and treatment effects.

Finally, I apply CoDiD to the effect of early-voting reform on the 2008 U.S.\
presidential election, comparing Maryland and New Jersey (treated) with
Pennsylvania and New York (control). The reform raised total turnout by about
$4.4\%$, increased the Democratic vote share by $0.92$ percentage points, and
reduced the Republican share by $1.18$ points, with the remaining mass
accruing to third parties. Additive level-DiD on the total is implausible in
this design: it would imply a turnout effect near zero purely because the
control bloc is roughly twice the size of the treated bloc.

\subsection*{Related literature}

This paper makes several methodological contributions at the intersection of difference-in-differences and compositional data analysis.

First, it contributes to the extensive DiD literature. Classical DiD focuses on scalar outcomes, estimating treatment effects on the mean, with surveys and applications summarized in \citet{DiD_Review2011, Review2, de2023two, roth2023s, baker2025difference}. Recent works have extended DiD to more complex settings: \citet{CallawaySant2020} develop methods for staggered treatment adoption, \citet{arkhangelsky2021synthetic} combine DiD with synthetic controls, and \citet{manski2018right, RamRoth2020, ban2022robust} relax the parallel trends assumption.

A second wave of research extends DiD to entire outcome distributions. Early contributions include quantile DiD \citep{meyer1990workers} and the changes-in-changes (CiC) approach \citep{athey2006identification}, which identifies counterfactual distributions nonparametrically. Subsequent work has generalized these ideas to multivariate outcomes \citep{torous2021optimal}, settings where identifying assumptions apply to cumulative distribution functions \citep{HavnesMogstad2015, RothSantanna2021}, copula-based approaches \citep{CallawayLiOka2018, CallawayLi2019, ghanem2023evaluating}, and methods using characteristic functions \citep{bonhommesauder2011}. However, these methods do not directly address categorical outcome distributions, which are shares.

\citet{graves2022difference} proposes a DiD for categorical outcomes while relying on linear parallel trends for proportions, which, as explained earlier, is generally inappropriate. \citet{zhou2025geodesic} extends the difference-in-differences framework to non-Euclidean data, including shares, by employing Fréchet means and geodesic transport. While their approach is mathematically elegant, the identifying assumptions are primarily geometric and offer limited economic interpretability. In contrast, my framework delivers both counterfactual quantities and shares that retain economic intuition while incorporating a geometric perspective specifically designed for categorical outcomes. Another related paper is \citet{UDID2024_Epi}, who propose a general DiD framework applicable to count data in the binomial setting. My analysis differs by focusing on the multinomial case, allowing for richer categorical structures and more general forms of distributional change.

Second, the paper contributes to the compositional data analysis (CoDA) literature. CoDA, pioneered by \citet{aitchison1982statistical, Aitchison1990, Aitchison1992, Aitchison2002}, provides tools for analyzing data constrained to the simplex. Subsequent developments \citep{egozcue2003isometric, BillheimerEtAl2001, BarceloVidalEtAl2001} refine transformations and models that respect the geometric structure of compositional data. More recently, \citet{arnold2020causal} connect CoDA to causal inference. I also combine compositional data methods with econometric identification strategies to analyze causal effects when outcomes are shares, providing a bridge between DiD and compositional statistics.

\subsection*{Outline}
\Cref{sec:setup} defines the setup and \Cref{sec:shares} develops CoDiD for the compositional margin. \Cref{sec:joint}
extends to joint identification of totals and shares. \Cref{sec:extensions}
treats multiple periods and partial identification. \Cref{sec:inf} develops
estimation and inference. \Cref{sec:app} presents the application.
\Cref{sec:conc} concludes. All proofs are in \Cref{app:proof1}.

\section{Setup and target parameters}\label{sec:setup}

There are two groups, $g\in\{0,1\}$, and two periods, $t\in\{0,1\}$. Group 1 receives treatment in period 1; group 0 remains untreated. Categories $k=1,\ldots,p$, with $p\geq2$, are mutually exclusive and exhaustive within a specified outcome universe. Category definitions and measurement conventions are held fixed across groups, periods, and treatment states.

\subsection{Where compositional outcomes come from}\label{sec:origins}

\paragraph{Distributions of categorical observations.}
Suppose that observation $i$ in cell $(g,t)$ has a category label $Y_{i,g,t}\in\{1,\ldots,p\}$. The labels may be unordered, as with political candidates or major choices. Let $Y_{g,t}$ denote the category of a randomly selected unit from the cell's population. Its distribution is
\[
\pi_{k,g,t}=\Pr(Y_{g,t}=k),\qquad \sum_k\pi_{k,g,t}=1.
\]
For $N_{g,t}$ observed units, let
\[
q_{k,g,t}=\sum_{i=1}^{N_{g,t}}\ind{Y_{i,g,t}=k},\qquad
\widehat\pi_{k,g,t}=\frac{q_{k,g,t}}{N_{g,t}}.
\]
Thus, the empirical distribution of categorical observations is itself a composition. 

\paragraph{Shares of quantities.}
Alternatively, the data may consist of a vector of quantities $q_{g,t}$ measured in common units: megawatt-hours generated by source, dollars spent by budget item, or enrollment recorded by institutional sector. Write $n_{g,t}=\sum_kq_{k,g,t}$. Its composition is $\pi_{g,t}=q_{g,t}/n_{g,t}$. The components need not be integers, and no individual-choice model is required. One can interpret $\pi_{k,g,t}$ as the probability that a unit of mass drawn uniformly from the total belongs to category $k$, but this is a representation of the quantity shares, not an assumption that individuals made discrete choices.

\paragraph{Implications.}
The two origins lead to the same mathematical object. A categorical distribution can be represented by the population masses assigned to its categories; a quantity vector yields a categorical distribution by normalization. The empirical interpretation is different, however: the sampling unit, the quantity unit, and the population whose size can respond to treatment must each be specified. The method applies to both origins, while the choice-model interpretation in \Cref{sec:choice} applies only when a behavioral model of category selection is appropriate.

\subsection{Potential quantities and the two margins}\label{sec:margins}

Let $q^d_{g,t}=(q^d_{1,g,t},\ldots,q^d_{p,g,t})'$ be the vector of potential population quantities under treatment status $d\in\{0,1\}$. These quantities may be population masses or expected counts. Realized sample counts are estimators of these objects, not automatically identical to them. Define
\[
 n^d_{g,t}=\sum_{k=1}^p q^d_{k,g,t},\qquad
 \pi^d_{g,t}=\frac{q^d_{g,t}}{n^d_{g,t}},\qquad
 \mathcal{S}^{p-1}_{++}=\{x\in\mathbb{R}^p:x_k>0,\ \one'x=1\}.
\]
The notation $n$ denotes the outcome total and need not be a fixed number of eligible individuals. In the voting application, it is votes cast, not the voting-eligible population. In the education application, it is enrollment recorded at covered institutions, not a resident cohort's college-going rate.

For a positive quantity vector, the representation $q^d_{g,t}=n^d_{g,t}\pi^d_{g,t}$ separates \emph{scale} from \emph{composition} without losing information. Knowing only $\pi^d_{g,t}$ loses scale: every positive multiple of $q^d_{g,t}$ has the same shares. Knowing only $n^d_{g,t}$ loses its allocation across categories. Consequently, a policy may expand all categories proportionally, redistribute a fixed total, or change both margins simultaneously. When the total is fixed by the definition of the population, only the second type of response is possible. Otherwise, separate identifying content is needed for the two margins.

\subsection{Treatment Effect Parameters}

\paragraph{Average Treatment Effect on the Treated (ATT).}
The ATT measures the causal change in the \emph{share} allocated to each
category:
\begin{equation}
\label{eq:att}
\mathrm{ATT}_k = \pi_{k,1,1}^1 - \pi_{k,1,1}^0
\quad (k = 1,\dots,p), \qquad
\mathrm{ATT} = \pi_{1,1}^1 - \pi_{1,1}^0 \in \bR^p .
\end{equation}
Since $\sum_k \pi_{k,1,1}^1 = \sum_k \pi_{k,1,1}^0 = 1$, the ATT
automatically satisfies $\sum_k \mathrm{ATT}_k = 0$: what one category gains,
the others collectively lose.

\paragraph{Compositional Treatment Effect on the Treated (CTT).}
The CTT uses the compositional difference operator (defined formally in
\Cref{sec:geom}) to capture the \emph{relative} redistribution of mass. Let
$r_k = \pi_{k,1,1}^1 / \pi_{k,1,1}^0$; then
\begin{equation}
\label{eq:ctt}
\mathrm{CTT} 
= \frac{1}{\sum_{j=1}^p r_j}\bigl(r_1, \dots, r_p\bigr).
\end{equation}
A component $\mathrm{CTT}_k > 1/p$ indicates that category $k$ gained
relative importance due to treatment. With no compositional effect,
$\mathrm{CTT} = (1/p, \dots, 1/p)$, the neutral element of the Aitchison
structure. The CTT is invariant to the overall scale of the effect, isolating
pure redistribution.

\paragraph{Functional Treatment Effect on the Treated (FTT).}
For any measurable $H\colon \cS^{p-1} \to \bR$,
\begin{equation}
\label{eq:ftt}
\mathrm{FTT}(H) = H(\pi_{1,1}^1) - H(\pi_{1,1}^0).
\end{equation}
A leading example is the Herfindahl--Hirschman index,
$H(\pi) = \sum_k \pi_k^2$. The correct procedure is to recover
$\pi_{1,1}^0$ via CoDiD and then compute $\mathrm{FTT}(\mathrm{HHI})$;
applying DiD directly to $\mathrm{HHI}(\pi_{g,t})$ ignores the simplex
constraint and conflates concentration effects with compositional
reallocation.

\section{Identification of the composition}\label{sec:shares}

\subsection{Parallel trends of log-odds and identification}

Choose category $p$ as the reference category. This choice is without loss of generality, as identification is invariant to the choice of reference category. For $k<p$, define the pairwise log-odds
\[
\ell_{k,g,t}^0=\log(\pi^0_{k,g,t}/\pi^0_{p,g,t}).
\]
The odds $\pi^0_{k,g,t}/\pi^0_{p,g,t}$ compare category $k$ with category $p$ within one group and period. I use this transformation as the basis for the restriction. I consider the following assumptions:

\begin{assumption}[No anticipation]\label{ass:no_ant}
The observed population quantities satisfy $q_{g,t}=q^0_{g,t}$ for $(g,t)\neq(1,1)$.
\end{assumption}

\begin{assumption}[Positivity]\label{ass:pos}
All relevant category quantities and probabilities are strictly positive.
\end{assumption}

\begin{assumption}[Parallel trends of log-odds (PTLO)]\label{ass:ptlo}
In the absence of treatment,
\begin{equation}\label{eq:ptlo}
 \ell^0_{k,1,1}-\ell^0_{k,1,0}
 =\ell^0_{k,0,1}-\ell^0_{k,0,0},\qquad k=1,\ldots,p-1.
\end{equation}
\end{assumption}

Parallel trends of log-odds (PTLO) requires equal changes in pairwise log-odds across groups, rather than equal changes in shares. Equivalently, the proportional change in the odds of $k$ versus $p$ must be the same in both groups. We can also see that PTLO is invariant to the reference category and does not impose an ordering on category labels. 

\begin{theorem}[Identification of counterfactual shares]\label{thm:shares}
Under \Cref{ass:no_ant,ass:pos,ass:ptlo},
\begin{equation}\label{eq:cf}
 \pi^0_{k,1,1}=\frac{\pi_{k,1,0}\pi_{k,0,1}/\pi_{k,0,0}}
 {\sum_{j=1}^p\pi_{j,1,0}\pi_{j,0,1}/\pi_{j,0,0}}.
\end{equation}
This is the unique composition satisfying \eqref{eq:ptlo}. Consequently, the share effects, the CTT, and any specified functional of the counterfactual composition are identified.
\end{theorem}

The formula is closed-form and guaranteed to produce a valid composition, so counterfactual shares outside $[0,1]$ cannot occur. The next two subsections show that the assumption has some very natural economic and geometric foundations.

\begin{remark}[Handling Zero Cells and Structural Zeros]
\label{rem:zeros}
\Cref{ass:pos} requires strict positivity. In practice, sampling zeros may occur in small cells, and structural zeros may arise from impossible categories. For sampling zeros, a principled approach is Laplace (Bayesian) smoothing: adding a small $\epsilon > 0$ to all counts before computing shares, which preserves the simplex and ensures the discrete copula is well-defined. For structural zeros, the discrete copula framework can be extended via \emph{I-projection} (information projection) onto the subspace of tables respecting the zero-pattern, as developed by \citet{geenens2020copula}. In our application, no structural zeros exist.
\end{remark}

\subsection{Discrete-copula interpretation}\label{sec:copula}

The discrete-copula interpretation answers the following question: what feature of the relationship between category and group membership remains unchanged under PTLO? Consider the case where the shares are generated by some individual choice data. 
In the DiD literature, \citet{ghanem2023evaluating} propose a copula stability assumption. This assumption imposes that the dependence between untreated outcomes for the treated and control groups in period~0 is the same as in period~1. When the outcomes are continuous, it is equivalent to the CiC model. However, for mixed or discrete distributions with ordered support, it yields only partial identification. I argue here that PTLO is similar in spirit for unordered discrete outcomes. 

What makes the copula invariance approach appealing is that classical copula theory provides an elegant separation between marginal distributions and the dependence structure of two continuous random variables, with the copula remaining invariant under any strictly increasing transformation of the variables. When the variables are discrete, however, this framework breaks down: the probability integral transform (PIT) no longer produces uniform random variables, and infinitely many copulas can be associated with the same discrete outcome. Moreover, the classical approach does not apply to unordered outcomes.

To separate marginal distributions from dependence in the discrete setting without encountering non-uniqueness, the notion of dependence structure must be redefined from a more fundamental perspective, one that does not rely on the PIT. This perspective, developed by \citet{geenens2020copula}, views a copula not as a particular function with uniform margins, but rather as an equivalence class of distributions sharing the same underlying dependence structure. The idea is simple. Consider the set of all bivariate probability mass functions (pmfs) defined on a given finite support. Define an equivalence relation on this set: two distributions are equivalent if one can be obtained from the other by independently rescaling its rows and columns. This operation, known as \emph{marginal distortion} or \emph{matrix scaling}, leaves the odds ratios of the probability table unchanged. Since odds ratios are margin-free measures of association, they provide a natural characterization of the dependence structure independently of the marginal distributions. Let $\Omega_{t}^0$ denote the odds-ratio matrix characterizing the dependence between $(Y_{t}^0, D)$.

\begin{proposition}[Discrete-copula interpretation of PTLO]
\label{prop:copula}
For strictly positive tables, the following statements are equivalent:
\begin{enumerate}[label=(\roman*)]
    \item Parallel trends in log-odds (PTLO) holds
    (\Cref{ass:ptlo}).
    \item $\Omega^{0}_{0}=\Omega^{0}_{1}$.
\end{enumerate}
\end{proposition}

PTLO therefore admits a discrete-copula interpretation: it requires the dependence structure between the untreated potential outcome and group membership to remain stable over time, while allowing their marginal distributions to evolve freely. In this sense, PTLO is a discrete-outcome analogue of the copula-stability restrictions used in continuous-outcome DiD and changes-in-changes models \citep{athey2006identification,ghanem2023evaluating}. The key restriction is on the dependence structure, rather than on the marginal distributions of outcomes or group membership.

\subsection{Economic Interpretation: A Structural Model of Choice}
\label{sec:choice}

Consider a population of participants in group $g$ at time $t$. Conditional on participating, individual $i$ chooses the alternative that maximizes
utility,
\[
U_{ik,g,t}^0 = V_{k,g,t}^0 + \varepsilon_{ik,g,t}^0 ,
\]
where $V_{k,g,t}^0$ is the systematic attractiveness of alternative $k$ and
$\varepsilon_{ik,g,t}^0 \sim$ i.i.d.\ Gumbel. The population share choosing
$k$ is the multinomial logit
\[
\pi_{k,g,t}^0 = \frac{\exp(V_{k,g,t}^0)}{\sum_{j=1}^p \exp(V_{j,g,t}^0)} .
\]
Since this section concerns only shares, participation is left unmodeled.
Define relative utilities
$\tilde{V}_{k,g,t}^0 = V_{k,g,t}^0 - V_{p,g,t}^0$ and pairwise differences
$\tilde{V}_{k,j,g,t}^0 = V_{k,g,t}^0 - V_{j,g,t}^0$, which capture the population's relative preference between alternatives $k$ and $j$.

\begin{proposition}[Equivalence, share margin]\label{prop:reduced}
Under the logit choice model, the following statements are equivalent:
\begin{enumerate}[label=(\roman*)]
    \item Parallel trends in log-odds (PTLO) holds
    (\Cref{ass:ptlo}).
    \item For all pairs $k,j$,
\[
\tilde{V}_{k,j,1,1}^0 - \tilde{V}_{k,j,1,0}^0
= \tilde{V}_{k,j,0,1}^0 - \tilde{V}_{k,j,0,0}^0 .
\]
\end{enumerate}
\end{proposition}

In a voting context, the odds of Democrat versus Republican quantify the population's relative
preference for Democrats; when they rise, the average attractiveness of
voting Democrat has increased relative to Republican. PTLO states
that, absent treatment, these relative preferences would have evolved in
parallel across groups for every pair of categories; any secular shift, such
as a national swing toward one party, would have affected both groups
identically. The treatment effect is the deviation from this common
trajectory.

\begin{remark}[Extensions to Mixed Logit and Semiparametric RUM]
\label{rem:mixed_rum}
\Cref{prop:reduced} establishes the equivalence between PTLO and parallel relative utilities under the standard multinomial logit model. A natural question is whether this extends to richer choice models. For \emph{mixed logit} (random coefficient) models, the aggregate share is an integral over the distribution of random coefficients. The equivalence continues to hold if the distribution of unobserved heterogeneity is strictly stationary across both groups and time. If the mixing distribution varies, parallel trends in systematic relative utilities no longer guarantee PTLO at the aggregate level. For \emph{semiparametric random-utility models}, the minimal behavioral condition for PTLO to retain its economic interpretation is that the latent index shifts are additively separable into group, time, and category effects, and that the joint distribution of unobserved preference shocks is invariant to group membership and time. Under these conditions, the relative choice probabilities evolve in parallel, preserving the discrete copula.
\end{remark}

\subsection{Geometric Interpretation: Aitchison Simplex Geometry}
\label{sec:geom}

The third reading requires neither the random-utility structure nor its
distributional assumptions. A natural counterfactual posits that the relative
rise or decline of each category in the control group is what would have
happened in the treated group absent treatment; equivalently, that the
counterfactual share path of the treated group is \emph{parallel} to the
control group's path, once ``parallel'' is read in the geometry intrinsic to
the simplex.

\begin{proposition}[\citealp{aitchison1982statistical,egozcue2003isometric}]
\label{pr1}
On the interior of $\cS^{p-1}$ define, for
$\pi^{1},\pi^{2}\in\operatorname{int}\cS^{p-1}$ and $a\in\bR$, the
perturbation and powering operators
\[
\pi^{1}\oplus\pi^{2}
= \frac{(\pi^{1}_{1}\pi^{2}_{1},\dots,\pi^{1}_{p}\pi^{2}_{p})}
       {\sum_{k}\pi^{1}_{k}\pi^{2}_{k}},
\qquad
a\odot\pi^{1}
= \frac{((\pi^{1}_{1})^{a},\dots,(\pi^{1}_{p})^{a})}
       {\sum_{k}(\pi^{1}_{k})^{a}} .
\]
Then $(\operatorname{int}\cS^{p-1},\oplus,\odot)$ is a real vector space of
dimension $p-1$, with $\oplus$ as addition, $\odot$ as scalar
multiplication, and the uniform composition $(1/p,\dots,1/p)$ as the zero
element.
\end{proposition}

The compositional difference is
$\pi^{1}\ominus\pi^{2}=\pi^{1}\oplus\bigl((-1)\odot\pi^{2}\bigr)$: the
perturbation carrying $\pi^{2}$ to $\pi^{1}$.

\begin{proposition}[Compositional difference and the CTT]\label{pr:comp}
Under \Cref{ass:ptlo}, the counterfactual composition satisfies
\[
\pi_{1,1}^{0}\ominus\pi_{1,0}^{0}=\pi_{0,1}^{0}\ominus\pi_{0,0}^{0}:
\]
the treated group's counterfactual displacement equals the control group's.
Under \Cref{ass:no_ant} in addition, the CTT is the Aitchison-space analogue of
the additive DiD contrast,
\begin{equation}
\label{codieq}
\mathrm{CTT}
= \bigl(\pi_{1,1}^{1}\ominus\pi_{1,0}^{0}\bigr)
  \ominus\bigl(\pi_{0,1}^{0}\ominus\pi_{0,0}^{0}\bigr).
\end{equation}
\end{proposition}

\Cref{figs} depicts the idea for $p=3$: the control path from
$\pi_{0,0}^{0}$ to $\pi_{0,1}^{0}$ and the treated counterfactual path from
$\pi_{1,0}^{0}$ to $\pi_{1,1}^{0}$ are equal perturbations, i.e.\ parallel
translations in the simplex.

\begin{figure}[tbp]
\centering
\pgfmathsetmacro{\lcI}{ln(.15/.30)}\pgfmathsetmacro{\lcII}{ln(.55/.30)}
\pgfmathsetmacro{\dI}{ln(.45/.30)-ln(.15/.30)}
\pgfmathsetmacro{\dII}{ln(.25/.30)-ln(.55/.30)}
\pgfmathsetmacro{\ltI}{ln(.50/.35)}\pgfmathsetmacro{\ltII}{ln(.15/.35)}
\begin{minipage}[t]{.48\textwidth}
\centering
{\small\textbf{(a) Category shares in the simplex}}\par\smallskip
\begin{tikzpicture}[x=.85cm,y=.85cm,font=\footnotesize]
  \coordinate (controlpre)  at ({sx(\lcI,\lcII)},{sy(\lcI,\lcII)});
  \coordinate (controlpost) at ({sx(\lcI+\dI,\lcII+\dII)},{sy(\lcI+\dI,\lcII+\dII)});
  \coordinate (treatedpre)  at ({sx(\ltI,\ltII)},{sy(\ltI,\ltII)});
  \coordinate (treatedcf)   at ({sx(\ltI+\dI,\ltII+\dII)},{sy(\ltI+\dI,\ltII+\dII)});
  \coordinate (additive) at ({3*.80+6*(-.15)},{3*sqrt(3)*.80});
  \fill[black!2] (0,0)--(6,0)--(3,{3*sqrt(3)})--cycle;
  \draw[black!60] (0,0)--(6,0)--(3,{3*sqrt(3)})--cycle;
  \node[below left]  at (0,0) {$\pi_3=1$};
  \node[below right] at (6,0) {$\pi_2=1$};
  \node[above] at (3,{3*sqrt(3)}) {$\pi_1=1$};
  \draw[red!35,thin,smooth] plot[domain=-2.6:3.2,samples=120,variable=\t]
     ({sx(\lcI+\t*\dI,\lcII+\t*\dII)},{sy(\lcI+\t*\dI,\lcII+\t*\dII)});
  \draw[blue!35,thin,smooth] plot[domain=-2.6:3.2,samples=120,variable=\t]
     ({sx(\ltI+\t*\dI,\ltII+\t*\dII)},{sy(\ltI+\t*\dI,\ltII+\t*\dII)});
  \draw[gray,dashed,-{Stealth[length=4pt]},shorten >=2.5pt] (controlpre)--(controlpost);
  \draw[gray,dashed,-{Stealth[length=4pt]}] (treatedpre)--(additive);
  \draw[red!75!black,thick,-{Stealth[length=5pt]},shorten >=2.5pt] plot[domain=0:1,samples=30,variable=\t]
     ({sx(\lcI+\t*\dI,\lcII+\t*\dII)},{sy(\lcI+\t*\dI,\lcII+\t*\dII)});
  \draw[blue!75!black,thick,-{Stealth[length=5pt]},shorten >=2.5pt] plot[domain=0:1,samples=30,variable=\t]
     ({sx(\ltI+\t*\dI,\ltII+\t*\dII)},{sy(\ltI+\t*\dI,\ltII+\t*\dII)});
  \fill[red!75!black]  (controlpre) circle (2pt);
  \fill[red!75!black]  (controlpost) circle (2pt);
  \fill[blue!75!black] (treatedpre) circle (2pt);
  \fill[blue!75!black] (treatedcf) circle (2pt);
  \draw[gray,thick] (additive) ++(-.07,-.07)--++(.14,.14);
  \draw[gray,thick] (additive) ++(-.07,.07)--++(.14,-.14);
  \node[above right,inner sep=2pt] at (controlpre)  {$\pi^0_{0,0}$};
  \node[right,inner sep=3pt]       at (controlpost) {$\pi^0_{0,1}$};
  \node[left,inner sep=3pt]        at (treatedpre)  {$\pi^0_{1,0}$};
  \node[above left,inner sep=2pt]  at (treatedcf)   {$\pi^0_{1,1}$};
  \node[above left,text=gray]      at (additive)    {$\widetilde\pi^{\mathrm{add}}_{1,1}$};
\end{tikzpicture}
\end{minipage}\hfill
\begin{minipage}[t]{.48\textwidth}
\centering
{\small\textbf{(b) The same paths in log-odds}}\par\smallskip
\begin{tikzpicture}
\begin{axis}[axis equal image,width=6.4cm,
 xmin=-1.05,xmax=1.9,ymin=-1.95,ymax=.95,
 xlabel={$\ell_1=\log(\pi_1/\pi_3)$},
 ylabel={$\ell_2=\log(\pi_2/\pi_3)$},
 label style={font=\footnotesize},tick label style={font=\scriptsize},
 xtick={-1,0,1},ytick={-1,0},grid=major,grid style={black!10},clip=false]
 \addplot[red!75!black,thick,-{Stealth[length=5pt]},shorten >=2.5pt,domain=0:1,samples=2]
   ({\lcI+x*\dI},{\lcII+x*\dII});
 \addplot[blue!75!black,thick,-{Stealth[length=5pt]},shorten >=2.5pt,domain=0:1,samples=2]
   ({\ltI+x*\dI},{\ltII+x*\dII});
 \addplot[red!75!black,only marks,mark=*,mark size=2pt]
   coordinates {(\lcI,\lcII) ({\lcI+\dI},{\lcII+\dII})};
 \addplot[blue!75!black,only marks,mark=*,mark size=2pt]
   coordinates {(\ltI,\ltII) ({\ltI+\dI},{\ltII+\dII})};
 \node[above right,font=\footnotesize,inner sep=2pt] at (axis cs:\lcI,\lcII) {$\ell^0_{0,0}$};
 \node[above right,font=\footnotesize,inner sep=2pt] at (axis cs:{\lcI+\dI},{\lcII+\dII}) {$\ell^0_{0,1}$};
 \node[above right,font=\footnotesize,inner sep=2pt] at (axis cs:\ltI,\ltII) {$\ell^0_{1,0}$};
 \node[above right,font=\footnotesize,inner sep=2pt] at (axis cs:{\ltI+\dI},{\ltII+\dII}) {$\ell^0_{1,1}$};
\end{axis}
\end{tikzpicture}
\end{minipage}
\par\smallskip
{\footnotesize\textcolor{red!75!black}{Red: comparison group}\qquad
\textcolor{blue!75!black}{Blue: treated group without treatment}\qquad
\tikz[baseline=-.5ex]\draw[gray,dashed,-{Stealth[length=3pt]}] (0,0)--(.6,0);\ Equal additive changes in shares}
\caption{ The red curve is the control group's
trajectory from pre-treatment ($\pi_{0,0}^{0}$) to post-treatment
($\pi_{0,1}^{0}$); the blue curve is the treated group's counterfactual
trajectory from $\pi_{1,0}^{0}$ to $\pi_{1,1}^{0}$. Dashed segments indicate
the common perturbation (parallel transport in Aitchison geometry).}\label{figs}
\end{figure}

\begin{remark}[Aitchison versus Euclidean parallelism]
The paths look curved in Euclidean coordinates because the simplex is a
nonlinear subset of $\bR^{p}$, but they are straight, parallel geodesics in
the Aitchison metric. Two Euclidean-parallel paths would generically leave
$\cS^{p-1}$ and yield an infeasible counterfactual; Aitchison-parallel paths
stay in the simplex by construction. This is the geometric reason the
counterfactual shares in \eqref{eq:cf} are always valid
probabilities.
\end{remark}

\section{Joint Identification of Totals and Shares}
\label{sec:joint}

The previous section addressed the compositional margin. In many
applications, the total is also of interest, and I introduce a corresponding
parameter.

\paragraph{Growth Treatment Effect on the Treated (GTT).}
The GTT measures the causal proportional change in the absolute size of each
category and of the total:
\begin{equation}
\label{eq:gtt}
\mathrm{GTT}_k = \frac{q_{k,1,1}^1}{q_{k,1,1}^0} - 1
\quad (k = 1,\dots,p), \qquad
\mathrm{GTT} = \frac{n_{1,1}^1}{n_{1,1}^0} - 1 .
\end{equation}
$\mathrm{GTT}_k = 0$ means treatment had no effect on the absolute quantity
in category $k$; $\mathrm{GTT}_k > 0$ means treatment caused that category to
grow.

\subsection{The log map}

Define $\varphi: \bR_{++}^p \to \bR^p$ by
$\varphi(q) = (\log q_1, \dots, \log q_p)$, a bijection.

\begin{assumption}[Parallel growth]\label{apg}
In the absence of treatment,
\[
\varphi(q_{1,1}^0) - \varphi(q_{1,0}^0)
= \varphi(q_{0,1}^0) - \varphi(q_{0,0}^0).
\]
\end{assumption}

\Cref{apg} states that, absent treatment, the proportional growth rate of
each category would have been the same in the two groups. It implies
\Cref{ass:ptlo} (by taking differences with category $p$) but is stronger, as it also restricts
the total.

\begin{theorem}[Joint identification]\label{thm:joint}
Under \Cref{ass:pos,ass:no_ant,apg}:  
\begin{enumerate}[label=(\roman*)]
\item the counterfactual quantities are
      $q_{k,1,1}^0 = q_{k,1,0} \, q_{k,0,1} / q_{k,0,0}$;
\item the counterfactual total is
      $n_{1,1}^0 = \sum_{k=1}^p q_{k,1,0}\, q_{k,0,1} / q_{k,0,0}$;
\item the counterfactual shares are
      $\pi_{k,1,1}^0 = q_{k,1,1}^0 / n_{1,1}^0$, and they coincide with
      those of \Cref{thm:shares}.
\end{enumerate}
\end{theorem}

\begin{remark}[Relation to ratio and Poisson DiD]\label{rem:poisson}
Category by category, the formula in (i) is the counterfactual implied by
``ratio'' parallel trends, as delivered by Poisson QMLE DiD
\citep{wooldridge2023simple} and by rate-based DiD in epidemiology
\citep{UDID2024_Epi}. The contribution of \Cref{thm:joint} is not the
per-category formula but the joint compositional structure: (a) the implied
counterfactual shares are automatically simplex-valid and coincide with the
shares identified from the strictly weaker odds ratio condition; (b) a single
assumption coherently identifies shares and total, with the decomposition
below making the wedge relative to log-total DiD explicit; (c) inference is
joint across categories, respecting cross-category dependence.
\end{remark}

\subsection{Implication for Totals: The Composition Adjustment Factor}

Parallel growth decomposes into a restriction on the total and a restriction on the composition:
\begin{equation}
\log n_{1,1}^0 - \log n_{1,0}^0
= \bigl(\log n_{0,1}^0 - \log n_{0,0}^0\bigr) + \log \lambda ,
\label{eq:decomp}
\end{equation}
where
\begin{equation}
\lambda = \sum_{k=1}^p \frac{\pi_{k,0,1}^0}{\pi_{k,0,0}^0}\, \pi_{k,1,0}^0
\label{eq:lambda}
\end{equation}
is a \emph{composition adjustment factor}: a weighted average of the control
group's share-growth ratios, weighted by the treated group's pre-treatment
shares. When the two groups share the same pre-treatment composition,
$\lambda = 1$ and parallel growth reduces to parallel trends in log totals;
otherwise $\log\lambda$ corrects the aggregation bias of log-total DiD.

To see why the adjustment matters, consider evaluating an early-voting
policy. Suppose the control state shows no change in total turnout, but its
vote composition shifts: the Democratic share rises while the Republican
share falls. Suppose the treated state's pre-treatment electorate is
disproportionately Republican, the group losing relative share in the
control state. Even with flat aggregate turnout in the control state, the
treated state's counterfactual turnout would have declined absent the policy,
because its composition loads on the shrinking group. Standard DiD in log
totals would wrongly impose no counterfactual change and thus mismeasure the
policy's effect.

\section{Extensions with Multiple Time Periods}
\label{sec:extensions}

\subsection{Pre-Trends Assessment and Testability}
\label{sec:pretrends_multi}

With $T_1 \geq 1$ pre-treatment periods beyond the reference period, the
credibility of \Cref{apg} can be assessed. Parallel growth in pre-treatment
periods requires the log-differences
$\Delta_{k,t} \equiv \log q_{k,1,t}^0 - \log q_{k,0,t}^0$ to be constant over $t \leq 0$ for each
$k$. A graphical
assessment appears in the application (\Cref{demfig}).

\begin{remark}[Event-study analogue and graphical assessment]
Plotting $\Delta_{k,t}$ against $t$ for each category provides a graphical argument: parallel pre-trends manifest as flat trajectories, and systematic deviations reveal the pattern and magnitude of any differential trends. This is particularly informative when
$T_1 \geq 2$.
\end{remark}

\subsection{Partial Identification under Robust Parallel Growth}
\label{sec:partial}

When pre-treatment log-differences are not exactly constant, I introduce a
relaxation in the spirit of \citet{ban2022robust}. Let $t \in \{-T_1, \dots, -1, 0, 1\}$, with
treatment at $t=1$. Define, for each category $k$,
\begin{equation}
\label{eq:dminmax}
d_k^{\min} = \min_{t \leq 0}\bigl(\log q_{k,1,t}^0 - \log q_{k,0,t}^0\bigr),
\qquad
d_k^{\max} = \max_{t \leq 0}\bigl(\log q_{k,1,t}^0 - \log q_{k,0,t}^0\bigr).
\end{equation}

\begin{assumption}[Robust parallel growth]\label{ra3}
For each category $k$, the post-treatment log-difference lies within the
range observed pre-treatment:
\begin{equation}
\label{eq:rpg}
d_{k,1} \equiv \log q_{k,1,1}^0 - \log q_{k,0,1}^0
\in \bigl[d_k^{\min},\; d_k^{\max}\bigr].
\end{equation}
\end{assumption}

When $d_k^{\min} = d_k^{\max}$ for all $k$, \Cref{apg} is recovered as a
special case; in general \Cref{ra3} is strictly weaker and delivers partial
identification.

\begin{theorem}[Sharp partial identification]\label{the2}
Under \Cref{ass:pos,ra3}, define
\begin{equation}
\label{eq:bounds_def}
b_k^{\min} = \exp\bigl(d_k^{\min}\bigr)\, q_{k,0,1}^0, \qquad
b_k^{\max} = \exp\bigl(d_k^{\max}\bigr)\, q_{k,0,1}^0 .
\end{equation}
The sharp identified set for the counterfactual quantity vector is the box
$\prod_{k=1}^{p}[b_k^{\min}, b_k^{\max}]$, and consequently
\begin{align}
q_{k,1,1}^0 &\in \bigl[b_k^{\min},\; b_k^{\max}\bigr], \label{eq:bound_q}\\
n_{1,1}^0 &\in \Bigl[\sum_{k=1}^p b_k^{\min},\;
                      \sum_{k=1}^p b_k^{\max}\Bigr], \label{eq:bound_S}\\
\pi_{1,1}^0 &\in \cP = \Bigl\{\pi \in \cS^{p-1}:
  \pi_k = q_k \big/ \textstyle\sum_j q_j,\;
  q_k \in [b_k^{\min}, b_k^{\max}] \;\forall k\Bigr\},
  \label{eq:bound_pi}
\end{align}
and the sets in \eqref{eq:bound_S}--\eqref{eq:bound_pi} are sharp for the
total and the share vector, respectively.
\end{theorem}

\begin{remark}[Estimation and inference for the bounds]
\label{rem:bounds_inf}
In practice, $d_k^{\min}$ and $d_k^{\max}$ are estimated by their sample
analogues, so the bounds themselves are random. Because each bound is a
smooth function (a minimum or maximum over finitely many periods of smooth
functions) of the cell log-counts, its sampling distribution follows from the
asymptotic theory of \Cref{sec:inf}, and confidence intervals for the
partially identified parameters can be constructed following
\citet{imbens2004confidence} and \citet{stoye2009more}. In the application,
sampling uncertainty is negligible relative to the width of the identified
set, so I report estimated bounds directly.
\end{remark}

\begin{remark}[Staggered treatment adoption]
\label{rem:staggered}
When units adopt treatment at different times, the $2\times 2$ result in
\Cref{thm:joint} generalizes by applying CoDiD within each cohort-specific
$2\times 2$ comparison, using never-treated or not-yet-treated units as
controls in the spirit of \citet{CallawaySant2020}, and aggregating
cohort-specific ATT and GTT parameters with the usual weights. A formal
treatment of staggered adoption for compositional outcomes is left for future
work.
\end{remark}

\section{Estimation and Inference}
\label{sec:inf}

\subsection{Sampling Model}
\label{sec:sampling}

For each cell $(g,t) \in \{0,1\}^2$, I observe $p$ counts drawn from a
multinomial distribution,
\begin{equation}
\label{eq:sampling}
(q_{1,g,t}, \dots, q_{p,g,t})
\;\sim\;
\mathrm{Multinomial}\bigl(n_{g,t};\; \pi_{1,g,t}^*, \dots,
\pi_{p,g,t}^*\bigr),
\end{equation}
where the cell size $n_{g,t}$ is fixed by design and $\pi_{k,g,t}^*$ are the
population probabilities. Counts are independent across cells. Let
$n = \sum_{g,t} n_{g,t}$.

\begin{assumption}[Regular sampling]\label{ass:samp}
As $n \to \infty$: (i) $n_{g,t}/n \to \rho_{g,t} \in (0,1)$ for all $(g,t)$,
with $\sum_{g,t}\rho_{g,t} = 1$; and (ii) $\pi_{k,g,t}^* \in (0,1)$ for all
$k$ and $(g,t)$, the sampling analogue of \Cref{ass:pos}.
\end{assumption}

Because the cell totals $n_{g,t}$ are fixed, all sampling variation comes
from the share vectors $\hat{\pi}_{g,t} = q_{g,t}/n_{g,t}$, and every CoDiD
estimator is a smooth function of the four independent vectors
$(\hat{\pi}_{0,0},\hat{\pi}_{0,1},\hat{\pi}_{1,0},\hat{\pi}_{1,1})$ and the
known cell sizes. I organize the asymptotics around \emph{centered
log-counts}, which are numerically identical to centered log-shares: writing
$q_{k,g,t}^{*} := n_{g,t}\pi_{k,g,t}^{*}$,
\begin{equation}
\label{eq:logcancel}
\log q_{k,g,t} - \log q_{k,g,t}^{*}
= \log\hat{\pi}_{k,g,t} - \log\pi_{k,g,t}^{*},
\end{equation}
since the $\log n_{g,t}$ terms cancel. This device keeps every delta-method
expansion anchored at a fixed point even though raw counts grow with $n$.

By the multivariate CLT for multinomial proportions,
$\sqrt{n_{g,t}}(\hat{\pi}_{g,t} - \pi_{g,t}^{*}) \xrightarrow{d}
\cN(\mathbf{0},\Sigma_{g,t})$ with
$\Sigma_{g,t} = \diag(\pi_{g,t}^{*}) - \pi_{g,t}^{*}(\pi_{g,t}^{*})^{\top}$,
and rescaling by the common rate $\sqrt{n}$ via \Cref{ass:samp} and Slutsky,
\begin{equation}
\label{eq:prop_clt_n}
\sqrt{n}\,(\hat{\pi}_{g,t} - \pi_{g,t}^{*})
\xrightarrow{d}
\cN\!\Bigl(\mathbf{0},\;\tfrac{1}{\rho_{g,t}}\Sigma_{g,t}\Bigr).
\end{equation}
The delta method applied to the componentwise log map, together with
\eqref{eq:logcancel}, yields
\begin{equation}
\label{eq:logcount_cell}
\sqrt{n}\,\bigl(\log q_{g,t} - \log q_{g,t}^{*}\bigr)
= \sqrt{n}\,\bigl(\log\hat{\pi}_{g,t} - \log\pi_{g,t}^{*}\bigr)
\xrightarrow{d}
\cN\!\Bigl(\mathbf{0},\;\tfrac{1}{\rho_{g,t}}A_{g,t}\Bigr),
\end{equation}
where
\begin{equation}
\label{eq:Agt}
A_{g,t}
= \diag\!\Bigl(\tfrac{1}{\pi_{g,t}^{*}}\Bigr)\,\Sigma_{g,t}\,
  \diag\!\Bigl(\tfrac{1}{\pi_{g,t}^{*}}\Bigr),
\qquad
[A_{g,t}]_{kj}
= \frac{\ind{k=j}}{\pi_{k,g,t}^{*}} - 1 .
\end{equation}

\paragraph{The counterfactual log-count vector.}
The plug-in estimator implied by \Cref{thm:joint} is
\begin{equation}
\label{eq:loghat}
\log\hat{q}_{1,1}^{0}
= \log q_{1,0} + \log q_{0,1} - \log q_{0,0},
\qquad
\hat{q}_{k,1,1}^{0} = \frac{q_{k,1,0}\,q_{k,0,1}}{q_{k,0,0}} .
\end{equation}
Since the true value satisfies
$\log q_{1,1}^{0*} = \log q_{1,0}^{*} + \log q_{0,1}^{*} -
\log q_{0,0}^{*}$, independence across cells gives
\begin{equation}
\label{eq:logcount_joint}
\sqrt{n}\,\bigl(\log\hat{q}_{1,1}^{0} - \log q_{1,1}^{0*}\bigr)
\xrightarrow{d}
\cN\!\bigl(\mathbf{0},\,\Sigma_c\bigr),
\qquad
\Sigma_c
= \frac{A_{1,0}}{\rho_{1,0}} + \frac{A_{0,1}}{\rho_{0,1}}
  + \frac{A_{0,0}}{\rho_{0,0}} .
\end{equation}
The treated post-period cell does not appear in $\Sigma_c$ because it does
not enter \eqref{eq:loghat}.

All estimators are smooth, scale-appropriate transformations of
\eqref{eq:logcount_joint} and of the observed cell $(1,1)$. Shares are
recovered through the softmax map
$\sigma(y)_k = e^{y_k}/\sum_j e^{y_j}$, which is invariant to adding a
constant to all coordinates; this shift-invariance is what allows the delta
method to be applied at the fixed point $\log\pi_{1,1}^{0*}$ even though
$\log q_{1,1}^{0*}$ drifts with $n$. Ratio-type parameters (the GTTs) are
exponentials of \emph{differences} of log-counts, which are likewise anchored
at fixed limits.

\paragraph{The parametric bootstrap.}

In practice, a parametric bootstrap is convenient for inference in this setting. For
$b = 1, \dots, B$, draw independently for each cell $(g,t)$
\[
(q_{1,g,t}^{(b)}, \dots, q_{p,g,t}^{(b)})
\sim \mathrm{Multinomial}\bigl(n_{g,t};\;\hat{\pi}_{1,g,t},
\dots,\hat{\pi}_{p,g,t}\bigr),
\]
recompute all estimators, and form percentile intervals from the empirical
$[\alpha/2,\,1-\alpha/2]$ quantiles of $\{\hat{\theta}^{(b)}\}_{b=1}^B$.
Because every estimator is a Hadamard-differentiable transformation of the
cell proportions, consistency of the parametric bootstrap should follow from
standard arguments \citep[Ch.~23]{vaart1998asymptotic}.

\section{Application: Early Voting and the 2008 U.S.\ Presidential Election}
\label{sec:app}

\subsection{Design}

Early voting allows registered voters to cast ballots before Election Day,
in person or by mail. I evaluate the causal effect of early-voting programs
on total turnout and on the compositional distribution of votes across
parties in the 2008 U.S.\ presidential election.

The treated group consists of Maryland and New Jersey, which introduced
early-voting programs before the 2008 election; Pennsylvania and New York
serve as the control group. The four states are geographically contiguous,
share comparable demographic and political profiles (\Cref{tabcomp}), and
consistently supported Democratic presidential candidates from 1992 to 2004,
providing a plausible basis for the parallel growth assumption.

\begin{figure}[h!]
\centering
\includegraphics[width=0.90\textwidth]{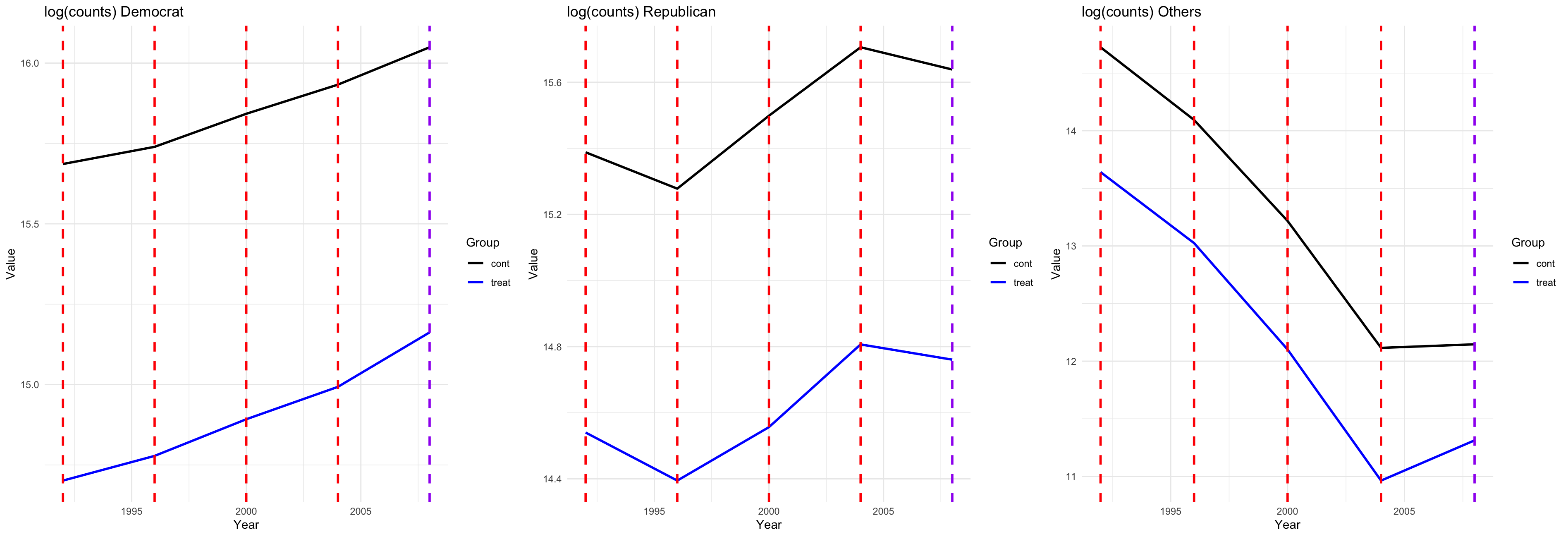}
\caption{Log-count evolution, 1992--2008 (treated vs.\ control). The dotted
red vertical line separates pre-treatment elections (1992--2004) from the
post-treatment election (2008). The broadly parallel pre-treatment
trajectories support the plausibility of parallel growth.}
\label{demfig}
\end{figure}

CoDiD requires two conditions. First, \emph{parallel growth} of log-vote
counts: \Cref{demfig} documents broadly parallel pre-treatment log-count
trajectories from 1992 to 2004 across all three party categories. Second,
\emph{no interference}: voters in one state are unlikely to change their party choice in response to early-voting laws in a neighboring state.

\begin{figure}[h!]
\centering
\includegraphics[width=0.90\textwidth]{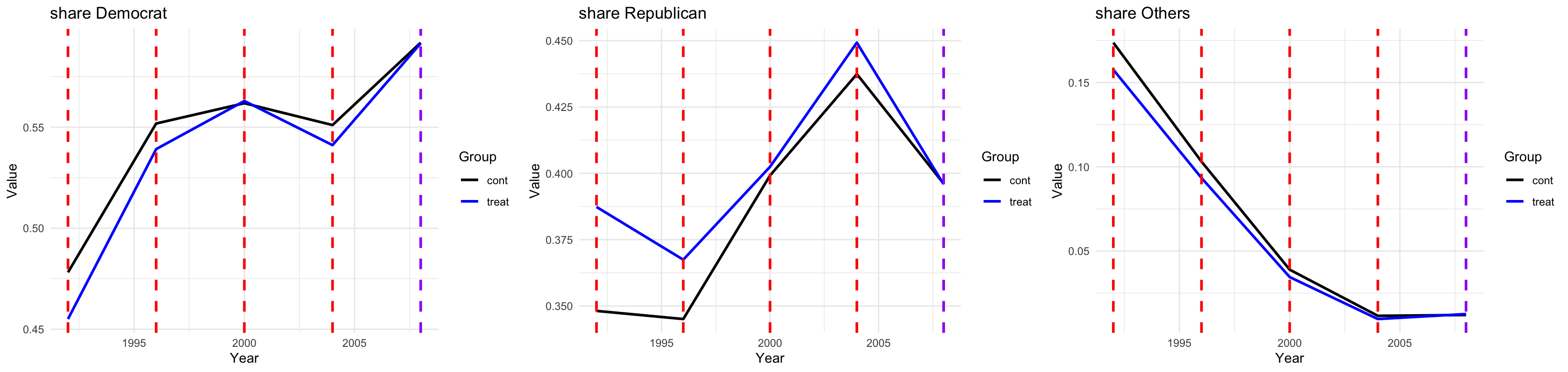}
\caption{Raw vote-share evolution, 1992--2008 (treated vs.\ control). The
non-parallel pre-treatment share trajectories illustrate why linear parallel
trends on shares is inappropriate.}
\label{shd}
\end{figure}

\begin{figure}[h!]
\centering
\includegraphics[width=0.90\textwidth]{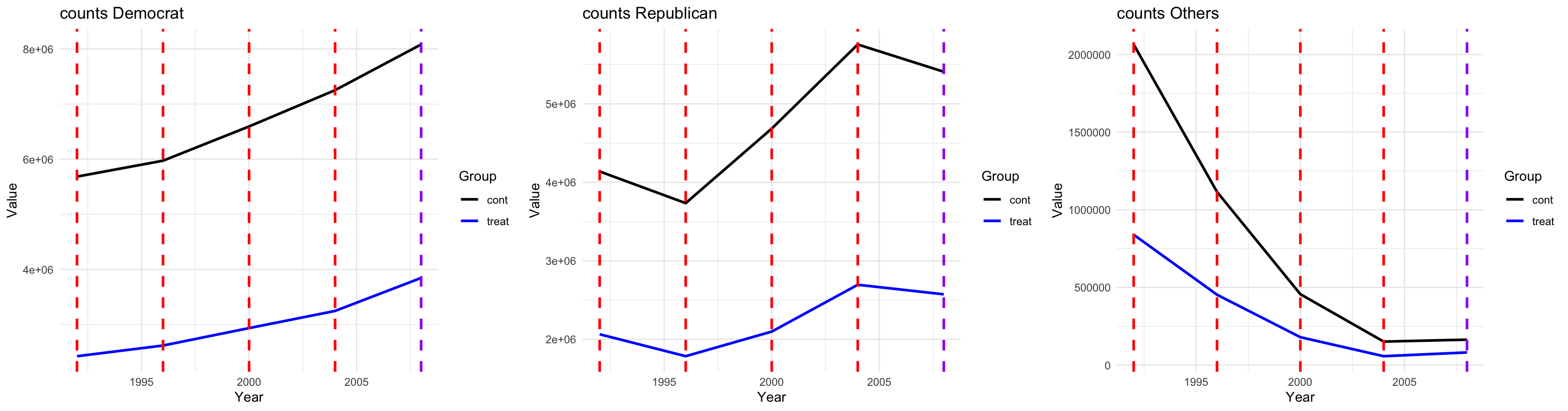}
\caption{Raw-count evolution, 1992--2008 (treated vs.\ control). The
large-scale difference between groups makes additive parallel trends on raw
quantities implausible.}
\label{coud}
\end{figure}

\paragraph{Why standard approaches may fail here.}
\Cref{shd} shows that raw vote shares do not exhibit parallel trends.
\Cref{coud} shows that the scale difference between groups (the control bloc
is roughly twice the size of the treated bloc) makes raw-quantity parallel
trends implausible: additive level-DiD on totals would attribute essentially
the whole control-group swing, in levels, to the much smaller treated bloc's
counterfactual. CoDiD's parallel growth assumption operates on
log-quantities, correcting both problems simultaneously.

\begin{table}[htbp]
\centering
\caption{Population Composition in 2007}
\label{tabcomp}
\begin{adjustbox}{max width=\textwidth}
\small
\setlength{\tabcolsep}{4pt}
\begin{tabular}{@{}lrrrrrrrr@{}}
\toprule
\textbf{State/Group} &
\textbf{Total Pop.} &
\textbf{White} &
\textbf{Black} &
\textbf{Am.\ Ind.} &
\textbf{Asian} &
\textbf{Two+} &
\textbf{Hisp./Latino} &
\textbf{White NH} \\
\midrule
Maryland     & 5,618,344  & 63.6\% & 29.5\% & 0.3\% & 5.0\%  & 1.6\% & 6.3\%  & 58.1\% \\
New Jersey   & 8,685,920  & 76.3\% & 14.5\% & 0.3\% & 7.5\%  & 1.3\% & 15.9\% & 62.2\% \\
\addlinespace
\textbf{MD+NJ} & 14,304,264 & 71.3\% & 20.8\% & 0.3\% & 6.5\% & 1.4\% & 12.1\% & 60.5\% \\
\addlinespace
New York     & 19,297,729 & 73.6\% & 17.3\% & 0.5\% & 6.9\%  & 1.5\% & 16.4\% & 60.3\% \\
Pennsylvania & 12,432,792 & 85.6\% & 10.8\% & 0.2\% & 2.4\%  & 1.0\% & 4.5\%  & 81.8\% \\
\addlinespace
\textbf{NY+PA} & 31,730,521 & 78.0\% & 14.5\% & 0.4\% & 5.0\% & 1.3\% & 11.7\% & 69.9\% \\
\bottomrule
\end{tabular}
\end{adjustbox}\\[4pt]
{\small\textit{Source: U.S.\ Census Bureau, Population Division, May 2008.}}
\normalsize
\end{table}

\subsection{Results}

\paragraph{Growth treatment effects.}
\Cref{tab2} reports estimated GTTs with $95\%$ parametric-bootstrap
confidence intervals ($B = 9{,}999$).

\begin{table}[h]
\centering
\caption{Estimated Growth Treatment Effects on the Treated (GTT)}
\label{tab2}
\begin{tabular}{lcc}
\toprule
\textbf{Party} & \textbf{Estimate} & \textbf{[95\% CI]} \\
\midrule
Democrat   & 0.0545 & [0.0533, 0.0557] \\
Republican & 0.0214 & [0.0199, 0.0230] \\
Other      & 0.3747 & [0.3578, 0.3915] \\
\midrule
\textbf{Total} & 0.0441 & [0.0441, 0.0443] \\
\bottomrule
\end{tabular}
\end{table}

The policy increased total turnout by approximately $4.4\%$, consistent with
the early-voting literature. All three categories experienced increases:
Democrats gained $5.5\%$, Republicans $2.1\%$, and third-party candidates
$37.5\%$ (the latter reflecting a small base). The differential Democratic
gain is consistent with evidence that early voting disproportionately
mobilizes left-leaning voters \citep{berry2025selective}.

\paragraph{Average treatment effects on shares.}
\Cref{tabshare1} reports the observed distribution, the counterfactual
distribution, and the ATT.

\begin{table}[h]
\centering
\caption{Estimated Categorical Distributions and ATT}
\label{tabshare1}

\begin{adjustbox}{max width=\textwidth}
\begin{tabular}{lccc}
\toprule
\textbf{Party} &
\textbf{Observed [CI]} &
\textbf{Counterfactual [CI]} &
\textbf{ATT [CI]} \\
\midrule
Democrat   & 0.5915 [0.5911, 0.5915] & 0.5823 [0.5818, 0.5829] &
             $+0.0092$ [0.0086, 0.0099] \\
Republican & 0.3959 [0.3954, 0.3962] & 0.4076 [0.4071, 0.4082] &
             $-0.0117$ [$-0.0124$, $-0.0111$] \\
Other      & 0.0126 [0.0125, 0.0127] & 0.0101 [0.0100, 0.0102] &
             $+0.0025$ [0.0024, 0.0027] \\
\bottomrule
\end{tabular}
\end{adjustbox}
\end{table}

Early voting increased the Democratic share by $0.92$ percentage points
(from a counterfactual of $58.23\%$ to an observed $59.15\%$), decreased the
Republican share by $1.17$ points, and increased the third-party share by
$0.25$ points. By construction, the ATT components sum to zero, confirming
internal consistency. These results are not recoverable from any collection
of separate binary DiDs: the counterfactual shares in \Cref{tabshare1} are
the unique feasible vector consistent with parallel growth.

\paragraph{Compositional treatment effects.}
\Cref{tabdem} reports the CTT, the \emph{relative} redistribution of mass.

\begin{table}[h]
\centering
\caption{Estimated Compositional Treatment Effect on the Treated (CTT)}
\label{tabdem}

\begin{tabular}{lcc}
\toprule
\textbf{Party} & \textbf{Estimate} & \textbf{[95\% CI]} \\
\midrule
Democrat   & 0.3136 & [0.3119, 0.3151] \\
Republican & 0.2998 & [0.2982, 0.3012] \\
Other      & 0.3867 & [0.3838, 0.3898] \\
\bottomrule
\end{tabular}
\end{table}

Each entry is the probability of supporting a given party under a
hypothetical equal baseline of $1/3$. Third-party support rises most relative
to the baseline ($38.7\%$ vs.\ $33.3\%$), while Republicans lose the most
relative importance ($30.0\%$). Both major parties lose relative ground to
third parties, but the ranking of the parties is unchanged by treatment.

\paragraph{What the confidence intervals do and do not measure.}
The multinomial sampling model treats each ballot as an independent draw
from a common cell-level distribution. With administrative data covering
millions of votes, the resulting sampling uncertainty is minuscule---hence
the extremely tight intervals in \Cref{tab2,tabshare1,tabdem}. These
intervals should be read as pure enumeration uncertainty, \emph{conditional
on the identifying assumption}. They do not capture uncertainty about
parallel growth itself, nor aggregate shocks common to voters within a
state, which in a two-versus-two-state design are absorbed into the
identifying assumption rather than the sampling distribution. 

\paragraph{Sensitivity: partial identification bounds.}
I apply \Cref{the2} using all pre-treatment elections (1992--2004) to
construct $[d_k^{\min}, d_k^{\max}]$. The resulting bounds on the ATT are
$[0.70,\, 1.20]$ percentage points for Democrats and $[-1.40,\, -0.90]$
percentage points for Republicans. The point estimates in \Cref{tabshare1}
($+0.92$ and $-1.17$ points) lie comfortably within the identified sets,
indicating robustness to deviations from exact parallel growth within the
range of historical pre-treatment variation.

\section{Conclusion}
\label{sec:conc}

This paper develops Compositional Difference-in-Differences, a unified
framework for causal inference on compositional vectors. Working first with
the compositional margin, parallel trends in log-odds identify the
counterfactual composition through a closed-form formula that always
produces valid shares, with a geometric interpretation as parallel transport
in Aitchison space and an economic interpretation as parallel trends in
relative preferences. Extending to log-counts jointly identifies effects on
the total and the composition from a single assumption, revealing a
composition adjustment factor that corrects the aggregation bias of
log-total DiD. The estimators are consistent and
asymptotically normal under multinomial sampling, pre-trends are testable,
and sharp bounds are available when parallel growth holds only approximately.
Natural extensions include staggered adoption, covariate adjustment through
conditional parallel growth, and synthetic-control analogues for
compositional outcomes.

\bibliographystyle{plainnat}
\bibliography{references}

\appendix
\section{Proofs}
\label{app:proof1}

\subsection{Proof of \texorpdfstring{\Cref{thm:shares}}{Theorem 1}}

Under Parallel Trends of Log-Odds (\Cref{ass:ptlo}), the odds ratios of the share tables are time-invariant. By the definition of the discrete copula and the subsequent discussion, this is equivalent to the odds ratios of the observable share tables being constant over time. Let category $p$ be the baseline category. For any category $k \in \{1, \dots, p-1\}$, the odds ratio of the treated group ($g=1$) relative to the control group ($g=0$) at time $t$ is:
\[
OR_{k,t} = \frac{\pi^0_{k,1,t} / \pi^0_{p,1,t}}{\pi^0_{k,0,t} / \pi^0_{p,0,t}}.
\]
\Cref{ass:ptlo} requires $OR_{k,0} = OR_{k,1}$ for all $k$. Thus:
\[
\frac{\pi^0_{k,1,0} / \pi^0_{p,1,0}}{\pi^0_{k,0,0} / \pi^0_{p,0,0}} = \frac{\pi^0_{k,1,1} / \pi^0_{p,1,1}}{\pi^0_{k,0,1} / \pi^0_{p,0,1}}.
\]
Rearranging to solve for the counterfactual share ratio relative to the baseline category yields:
\[
\frac{\pi^0_{k,1,1}}{\pi^0_{p,1,1}} = \frac{\pi_{k,1,0}}{\pi_{p,1,0}} \frac{\pi_{k,0,1} / \pi_{k,0,0}}{\pi_{p,0,1} / \pi_{p,0,0}},
\]
where I use the fact that pre-treatment and control shares are observed. Let $W_k = \pi_{k,1,0} \pi_{k,0,1} / \pi_{k,0,0}$ for all $k=1,\dots,p$. The equation above implies that $\pi^0_{k,1,1} / \pi^0_{p,1,1} = W_k / W_p$. Because shares must sum to one, normalizing these ratios gives the closed-form solution:
\begin{equation}
\label{eq:cf_shares}
\pi^0_{k,1,1} = \frac{W_k}{\sum_{j=1}^p W_j} = \frac{\pi_{k,1,0} \pi_{k,0,1} / \pi_{k,0,0}}{\sum_{j=1}^p \pi_{j,1,0} \pi_{j,0,1} / \pi_{j,0,0}}.
\end{equation}
Under \Cref{ass:pos}, all observed shares are strictly positive, so $W_k > 0$ for all $k$. This guarantees the denominator is strictly positive and $\pi^0_{k,1,1} \in (0,1)$, ensuring the counterfactual lies strictly in the interior of the simplex. \hfill$\square$

\subsection{Proof of \texorpdfstring{\Cref{prop:reduced}}{Proposition 2}}

Under the multinomial logit model, the share of category $k$ is $\pi^0_{k,g,t} = \frac{\exp(V^0_{k,g,t})}{\sum_j \exp(V^0_{j,g,t})}$. The ratio of shares between category $k$ and the baseline category $p$ is:
\[
\frac{\pi^0_{k,g,t}}{\pi^0_{p,g,t}} = \exp(V^0_{k,g,t} - V^0_{p,g,t}) = \exp(\tilde{V}^0_{k,g,t}).
\]
The odds ratio $OR_{k,t}$ defined in the proof of \Cref{thm:shares} can be rewritten as:
\[
OR_{k,t} = \frac{\pi^0_{k,1,t} / \pi^0_{p,1,t}}{\pi^0_{k,0,t} / \pi^0_{p,0,t}} = \frac{\exp(\tilde{V}^0_{k,1,t})}{\exp(\tilde{V}^0_{k,0,t})} = \exp(\tilde{V}^0_{k,1,t} - \tilde{V}^0_{k,0,t}).
\]
\Cref{ass:ptlo} requires $OR_{k,0} = OR_{k,1}$ for all $k$. Taking the natural logarithm of both sides gives:
\[
\tilde{V}^0_{k,1,0} - \tilde{V}^0_{k,0,0} = \tilde{V}^0_{k,1,1} - \tilde{V}^0_{k,0,1},
\]
which rearranges to $\tilde{V}^0_{k,1,1} - \tilde{V}^0_{k,1,0} = \tilde{V}^0_{k,0,1} - \tilde{V}^0_{k,0,0}$. This is exactly the parallel trend of relative utilities with respect to the baseline category $p$.
Since $\tilde{V}^0_{k,j,g,t} = (V^0_{k,g,t} - V^0_{p,g,t}) - (V^0_{j,g,t} - V^0_{p,g,t}) = \tilde{V}^0_{k,g,t} - \tilde{V}^0_{j,g,t}$, the parallel trends condition holds for the baseline $p$ if and only if it holds for all pairs $(k,j)$. \hfill$\square$

\subsection{Proof of \texorpdfstring{\Cref{pr:comp}}{Proposition 4}}

Under \Cref{ass:ptlo}, I established that $OR_{k,0} = OR_{k,1}$ for all $k$. This implies:
\[
\frac{\pi^0_{k,1,1} / \pi^0_{k,1,0}}{\pi^0_{p,1,1} / \pi^0_{p,1,0}} = \frac{\pi^0_{k,0,1} / \pi^0_{k,0,0}}{\pi^0_{p,0,1} / \pi^0_{p,0,0}}.
\]
This means the vector of relative growth ratios for the treated group, $\mathbf{r}^1 = (\pi^0_{k,1,1} / \pi^0_{k,1,0})_{k=1}^p$, and the control group, $\mathbf{r}^0 = (\pi^0_{k,0,1} / \pi^0_{k,0,0})_{k=1}^p$, are proportional to each other: $r^1_k = C \cdot r^0_k$ for some constant $C > 0$.
By \Cref{pr1}, the compositional difference $\pi^1 \ominus \pi^2$ is the perturbation carrying $\pi^2$ to $\pi^1$, which has components proportional to $\pi^1_k / \pi^2_k$, normalized to sum to 1. Since $\mathbf{r}^1$ and $\mathbf{r}^0$ are proportional, normalizing them to sum to 1 yields the exact same composition. Therefore:
\[
\pi^0_{1,1} \ominus \pi^0_{1,0} = \pi^0_{0,1} \ominus \pi^0_{0,0}.
\]
For the CTT formula, I use the fact that $(\operatorname{int}\cS^{p-1}, \oplus)$ is an abelian group, which implies $a \ominus b = (a \ominus c) \ominus (b \ominus c)$ for any $c$. Letting $a = \pi^1_{1,1}$, $b = \pi^0_{1,1}$, and $c = \pi^0_{1,0}$, and using the no-anticipation assumption \Cref{ass:no_ant} $\pi^0_{1,0} = \pi^1_{1,0}$, I get:
\[
\mathrm{CTT} = \pi^1_{1,1} \ominus \pi^0_{1,1} = \bigl(\pi^1_{1,1} \ominus \pi^0_{1,0}\bigr) \ominus \bigl(\pi^0_{1,1} \ominus \pi^0_{1,0}\bigr).
\]
Substituting the parallel transport result $\pi^0_{1,1} \ominus \pi^0_{1,0} = \pi^0_{0,1} \ominus \pi^0_{0,0}$ yields \eqref{codieq}. \hfill$\square$

\subsection{Proof of \texorpdfstring{\Cref{thm:joint}}{Theorem 2}}

Under Parallel Growth (\Cref{apg}), for each category $k$:
\[
\log q^0_{k,1,1} - \log q^0_{k,1,0} = \log q^0_{k,0,1} - \log q^0_{k,0,0}.
\]
By \Cref{ass:no_ant}, $q^0_{k,1,0} = q_{k,1,0}$ is observed. The control group is never treated, so $q^0_{k,0,1} = q_{k,0,1}$ and $q^0_{k,0,0} = q_{k,0,0}$ are observed. Rearranging the log equation and exponentiating gives:
\[
q^0_{k,1,1} = \frac{q_{k,1,0} q_{k,0,1}}{q_{k,0,0}},
\]
proving (i). Summing over $k$ yields the counterfactual total $n^0_{1,1} = \sum_k q^0_{k,1,1}$, proving (ii).
For (iii), the counterfactual shares are $\pi^0_{k,1,1} = \frac{q^0_{k,1,1}}{n^0_{1,1}}$. The ratio of counterfactual shares between category $k$ and baseline $p$ is:
\[
\frac{\pi^0_{k,1,1}}{\pi^0_{p,1,1}} = \frac{q^0_{k,1,1}}{q^0_{p,1,1}} = \frac{q_{k,1,0} q_{k,0,1} / q_{k,0,0}}{q_{p,1,0} q_{p,0,1} / q_{p,0,0}}.
\]
Since $q_{k,g,t} = n_{g,t} \pi_{k,g,t}$, the total $n_{g,t}$ cancels out in the ratio:
\[
\frac{\pi^0_{k,1,1}}{\pi^0_{p,1,1}} = \frac{\pi_{k,1,0} \pi_{k,0,1} / \pi_{k,0,0}}{\pi_{p,1,0} \pi_{p,0,1} / \pi_{p,0,0}}.
\]
This is exactly the relation derived from PTLO in the proof of \Cref{thm:shares}. Since the ratios to the baseline category uniquely determine the simplex vector (via normalization), the shares identified by Parallel Growth coincide exactly with those identified by PTLO. \hfill$\square$

\subsection{Derivation of the Decomposition \texorpdfstring{\eqref{eq:decomp}}{(9)}}

From part (ii) of \Cref{thm:joint}, substituting $q_{k,g,t} = n^0_{g,t} \pi^0_{k,g,t}$:
\begin{align*}
n^0_{1,1} &= \sum_{k=1}^p q^0_{k,1,1} = \sum_{k=1}^p q_{k,1,0} \frac{q_{k,0,1}}{q_{k,0,0}} \\
&= \sum_{k=1}^p \left( n_{1,0} \pi_{k,1,0} \right) \frac{n_{0,1} \pi_{k,0,1}}{n_{0,0} \pi_{k,0,0}} \\
&= n_{1,0} \frac{n_{0,1}}{n_{0,0}} \sum_{k=1}^p \pi_{k,1,0} \frac{\pi_{k,0,1}}{\pi_{k,0,0}}.
\end{align*}
Taking the natural logarithm of both sides yields $\log n^0_{1,1} - \log n_{1,0} = (\log n_{0,1} - \log n_{0,0}) + \log \lambda$, where $\lambda = \sum_{k=1}^p \pi_{k,1,0} \frac{\pi_{k,0,1}}{\pi_{k,0,0}}$, which is exactly \eqref{eq:decomp} and \eqref{eq:lambda}. \hfill$\square$

\subsection{Proof of \texorpdfstring{\Cref{the2}}{Theorem 3}}

Under \Cref{ass:pos}, all observed quantities are strictly positive. \Cref{ra3} states that the counterfactual log-difference $d_{k,1} = \log q^0_{k,1,1} - \log q_{k,0,1}$ lies in the interval $[d_k^{\min}, d_k^{\max}]$. Exponentiating, I get $q^0_{k,1,1} \in [\exp(d_k^{\min}) q_{k,0,1}, \exp(d_k^{\max}) q_{k,0,1}] = [b_k^{\min}, b_k^{\max}]$.
Since the bounds apply independently to each category, the sharp identified set for the counterfactual vector $q^0_{1,1}$ is the Cartesian product (box) $\mathcal{B} = \prod_k [b_k^{\min}, b_k^{\max}]$.

\emph{Sharpness of the box.} For any $q \in \mathcal{B}$, I can construct a valid data-generating process by setting $q^0_{k,1,1} = q_k$. The implied log-differences $d_{k,1} = \log q_k - \log q_{k,0,1}$ fall exactly within $[d_k^{\min}, d_k^{\max}]$, satisfying \Cref{ra3}. Since the observed data place no restrictions on $q^0_{1,1}$ other than \Cref{ra3}, every point in $\mathcal{B}$ is attainable, making $\mathcal{B}$ the sharp identified set for $q^0_{1,1}$.

\emph{Totals and shares.} The identified set for any continuous functional of $q^0_{1,1}$ is the image of $\mathcal{B}$ under that functional. The total $n^0_{1,1} = \sum_k q_k$ is continuous and strictly increasing in each argument, so its minimum and maximum over the compact box $\mathcal{B}$ are attained at the lower and upper corners, yielding \eqref{eq:bound_S}.
The share map $\pi_k = q_k / \sum_j q_j$ is continuous on $\mathcal{B} \subset \bR_{++}^p$. Its image is exactly the set $\cP$ defined in \eqref{eq:bound_pi}. Because every $q \in \mathcal{B}$ is attainable, every $\pi \in \cP$ is attainable, making $\cP$ the sharp identified set for the shares. \hfill$\square$

\end{document}